\documentclass{pasj00}

\begin{document}
\SetRunningHead{S. Takahashi and K. Yamashita}{XMM-Newton Observation of
 Abell 1650}
\Received{2003/07/31}
\Accepted{2003/09/29}

\title{XMM$-$Newton Observation of the Cluster of Galaxies Abell 1650}

\author{Seiji \textsc{Takahashi} and Koujun \textsc{Yamashita}}
\affil{Department of Physics, Nagoya University, Furo-cho, Chikusa-ku,
Nagoya 464-8602}
\email{takahasi@u.phys.nagoya-u.ac.jp}


\maketitle


\begin{abstract} 

We present the nonuniform distribution of temperature and abundance in
 the angular extent of $10\arcmin$ (1.4 Mpc) in A~1650 observed by {\rm
 XMM-Newton}. Spectral analysis was carried out in 19
 systematically-subdivided regions, and in 15 specified regions referred
 to the hardness ratio ($HR$) map. The value of $HR$, defined as the
 ratio of counts in the 1.6--10 keV band to those in the 0.8--1.6 keV,
 is an indicator to investigate the spatial variation of spectral
 features. The temperature and abundance were obtained to be 4--6 keV
 and 0.2--0.5 solar value with average values of $5.62^{+0.05}_{-0.07}$
 keV  and $0.36^{+0.02}_{-0.01}$ solar value within a radius of
 $5\arcmin$, respectively. The redshift was derived to be
 $8.01^{+0.02}_{-0.02}\times 10^{-2}$ against an optical value of
 $8.45\times 10^{-2}$. It turned out that cool regions with a
 temperature of around 4 keV and a scalelength of a few 100 kpc are
 patchily distributed in the outer envelope. This temperature structure
 could be explained by the infalling of groups of galaxies into the
 cluster main body. More than 30 point sources were found around A~1650
 in the FOV of $30\arcmin$. The spectrum of the Galactic diffuse
 emission was obtained in the outskirts of A~1650. 
\end{abstract}


\section{Introduction}

The structure and evolution of clusters of galaxies are considered in
the scenario of gravitational infalling of groups of galaxies and the
merging of clusters. The process may be described as a sequence of
merging and relaxation phases in which substructures are alternatively
accreted and merged into the larger system. {\it N}-body hydrodynamic
simulations show that the merger of major cluster components produces
pronounced nonuniformities of the plasma temperature, gas density, and
magnetic field strength in an intracluster medium (ICM) (\cite{roe97};
\cite{roe00}; \cite{ric98}). Therefore, the temperature distribution in
the ICM is sensitive to the processes of hierarchical clustering.

Previously, {\rm ASCA}, {\rm ROSAT}, and {\rm BeppoSAX} demonstrated
that many clusters have complex temperature structures in an ICM,
suggesting that clusters have experienced a recent merger and are still
forming (\cite{mar98}; \cite{fur01}; \cite{wat99}). Recent X-ray
observations of {\rm XMM-Newton} and {\rm Chandra} enable us to study
spatially resolved ICM X-ray spectra with excellent angular resolution
and high sensitivity. The first observations show that the central
regions of clusters are not simple spherically-symmetric systems, but
are morphologically and spectroscopically very complex (e.g., A~2142:
\cite{mar00}; A~1795: \cite{tam01}; S${\rm \acute{e}}$rsic 159-03:
\cite{kaa01}; Centaurus: \cite{san02}).

In this paper, we present the first {\rm XMM-Newton} EPIC observation of
the cluster of galaxies A~1650, which is characterized as a richness
class 2, Bautz--Morgan type I--II and optical redshift of $z$ = 0.0845
(\cite{abe89}; \cite{str99}). Previous X-ray observations with {\rm
Einstein} \citep{whi97}, {\rm ROSAT} \citep{per98}, and {\rm ASCA}
(\cite{ike02}; \cite{whi00}) have shown that A~1650 has an average
temperature of approximately 6 keV, metallicities close to 0.3, an X-ray
luminosity of $7.9 \times 10^{44}$ erg s$^{-1}$ in the $0.1$--$2.4$ keV
band, contains a moderate cooling flow with a cooling rate of $\dot{M} =
280_{-80}^{+464}~M_{\odot}~{\rm yr}^{-1}$ and a cD galaxy 2MASXi\,
J1258414$-$014541 in the center, which is radio quiet \citep{bag94}.

Throughout this paper we assume $H_0 = 50~{\rm km~ s}^{-1}~{\rm
Mpc}^{-1}$. One arcminute corresponds to 144 kpc at a redshift of
0.0845. We use the abundance measurements of \citet{and89} for spectral
fitting, in which Fe/H $= 4.68\times10^{-5}$ by number. Errors are
quoted at the 90\% confidence level, unless otherwise stated.


\section{Observation and Data Reduction}

A~1650, located at
$\alpha=\timeform{12h58m41.5s},~\delta=\timeform{-01D45'41.4''}~[{\rm
J}2000]$ (corresponding to galactic coordinates $l=306\fdg 68,~b=61\fdg
06$), was observed for 43 ks by the {\rm XMM-Newton} satellite
\citep{jan01} in revolution 377 on  2001 December 29--30. The three EPIC
cameras, MOS1, MOS2, and PN (\cite{tur01}; \cite{str01}) were operated
in the full window mode with a medium filter.

We generated qualified event files using the standard tasks {\sc emchain}
and {\sc epchain} in the XMM SAS version 5.3.3 software. X-ray events
with patterns 0 to 12 for the two MOS cameras were used, whereas for the
PN single and double-pixel events were selected.

The light curve of this observation shows a partially and unusually high
count rate due to the background induced by solar flares. During flares,
the observed spectrum shows a hard continuum easily detectable above 10
keV. Therefore, we extracted the light curves over the whole field of
view in the energy range of 10 to 15 keV, and then excluded all time
intervals during solar flares.
The selected events lead to effective exposure times of 37.1 ks, 38.1
ks, and 31.0 ks for the MOS1, MOS2, and PN cameras, respectively.

Blank-sky data sets were prepared from several pointings at high galactic
latitude by removing point sources, and were used as background for all
of this analysis. We applied the same screening criteria for the
blank-sky backgrounds, which include the particle and sky
background. CLOSED filter data were used for the particle background,
and were normalized to derive the sky background while referring to the
count rate in the energy band of 10 to 12 keV for MOSs and 12 to 14 keV
for PN, because the particle background level is variable at about
10\%. With this method, we corrected for the particle background. We
also took into account the soft cosmic X-ray background (CXB), which
depends on the viewing direction of the sky. To estimate the soft
CXB, we used regions of the detector in which X-ray emission from the
cluster was negligible. We extracted a spectrum in the annulus between
$10\arcmin$ and $15\arcmin$ from the cluster center, excluding bright
point sources, and derived the normalization factor between the
blank-sky fields and A~1650 observations. The CXB in blank-sky fields
and around A~1650 in the soft band (0.5--3.0 keV) differ by less than
3\%. We thus used the above normalization factor. Background spectra
have been accumulated from the same detector areas as the source spectra.

Finally, a vignetting correction was applied to the event files by
using a method proposed by \citet{arn01}.
Bright streak pixels for both MOSs and PN, and the pixels with high
electric noise close to CCD border for PN were masked out manually.


\section{Cluster Morphology}

\begin{figure}
  \begin{center}
    \FigureFile(80mm,80mm){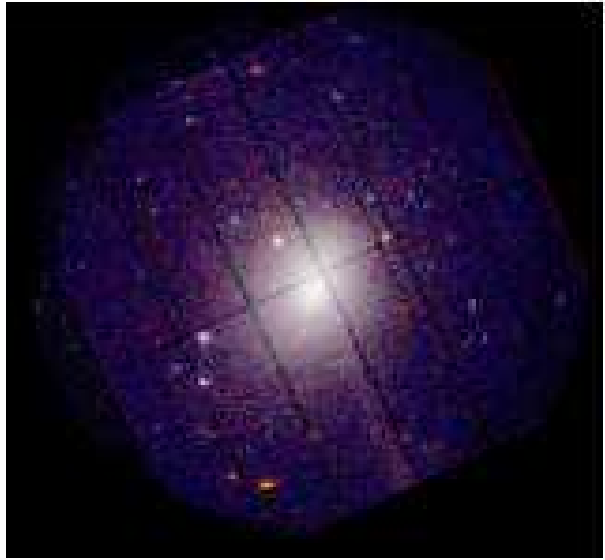}
  \end{center}
  \caption{X-ray color image combining MOSs and PN of the central
 30$\arcmin \times 30\arcmin$ of A~1650. The red, green, and
 blue intensities correspond to counts from energy bands of 0.3--1.3 keV,
 1.3--2.0 keV, and 2.0--10.0 keV respectively. The lines are due to the
 gaps between CCD chips.}\label{fig:XMM_color}
\end{figure}

\begin{figure}
  \begin{center}
    \FigureFile(80mm,80mm){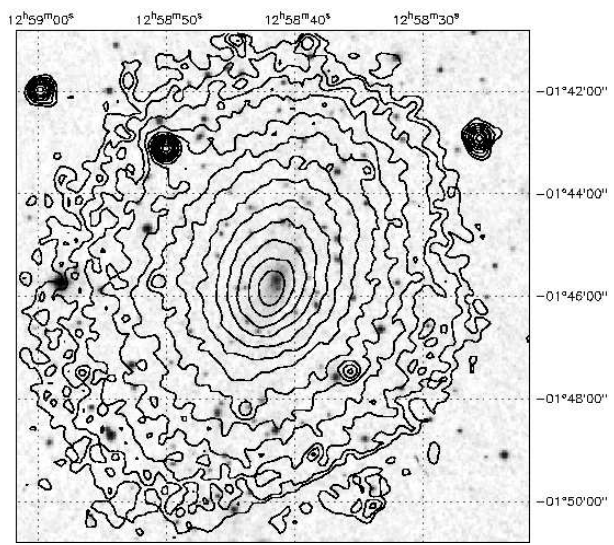}
  \end{center}
  \caption{MOSs X-ray surface brightness contours logarithmically
 spaced by a factor of $\sqrt{2}$, binned by $4\farcs4$ pixels and
 filtered with a $\sigma=4\farcs4$ Gaussian in the 0.8--10 keV band
 after subtracting background, overlaid on the Digitized Sky Survey
 optical image. The field is $10\arcmin\times 10\arcmin$ with north up
 and east to the left. The linear artifacts are due to the
 gaps between CCD chips.}\label{fig:mos_opt}
\end{figure}

The whole X-ray image in the field of view was constructed in color-coded
energy bands of 0.3--1.3 keV (red), 1.3--2 keV (green), and 2--10 keV
(blue) by combining MOSs and PN event files with a pixel size of
$4\farcs4$, as shown in figure \ref{fig:XMM_color}. The effects of
out-of-time events can be seen as a thin luminous band in the lower
chip, fourth from the left.

The X-ray color image represents the surface brightness distribution and
spectral feature. Soft sources appear red, moderately hard sources
appear green, and the hardest sources appear blue. The angular size of
A~1650 seems to extend to a radius of $10\arcmin$. Therefore, we show an
X-ray surface brightness distribution in the 0.8--10 keV band from the
two weighted MOS event files with a pixel size of $4\farcs4$ and a
Gaussian filter with $\sigma=4\farcs4$ in a $10\arcmin\times 10\arcmin$
region centered on the brightness peak and overlaid on the DSS optical
image in figure \ref{fig:mos_opt}. The X-ray morphology of A~1650 is
centrally concentrated and slightly elongated with an elliptical shape
and a position angle of $\sim 20\arcdeg$ (north to west). The optical
center of the cD galaxy and the peak position of the X-ray surface
brightness are displaced by about $10\arcsec$ (23 kpc) toward the
southeast. The orientation of cD galaxy is well-aligned to that of the
cluster.

There is no clear evidence of substructures, suggesting that the cluster
has not experienced a major merger in the recent past, and is in a
relaxed state. To identify significant substructures in the core, we
fitted the combined MOSs image with a 2D elliptical $\beta$-model and
built up a map of the residuals of the data over the best-fit
model. Images were extracted in the 0.8--10 keV band from the weighted
MOS event files in a pixel of size $4\farcs4$, and added to make a
combined MOS image. We fit the image within an angular radius of
$8\arcmin$, excluding bright point sources and the region close to the
CCD border. We fixed the centroid parameters to the X-ray emission peak,
and allowed the core radius $r_c$, index $\beta$, ellipticity
$\epsilon$, position angle $PA$ and normalization $n_0$ to vary. The
best-fit elliptical $\beta$-model, $n = n_0[1+(r/r_c)^2]^{-3\beta/2}$,
yields a core radius, $r_c$, of $0\farcm86$ (123 kpc), a $\beta$ of
0.57, an $\epsilon$ of 0.26, and a $PA$ of $-22\fdg0$.

\begin{figure}
  \begin{center}
    \FigureFile(80mm,80mm){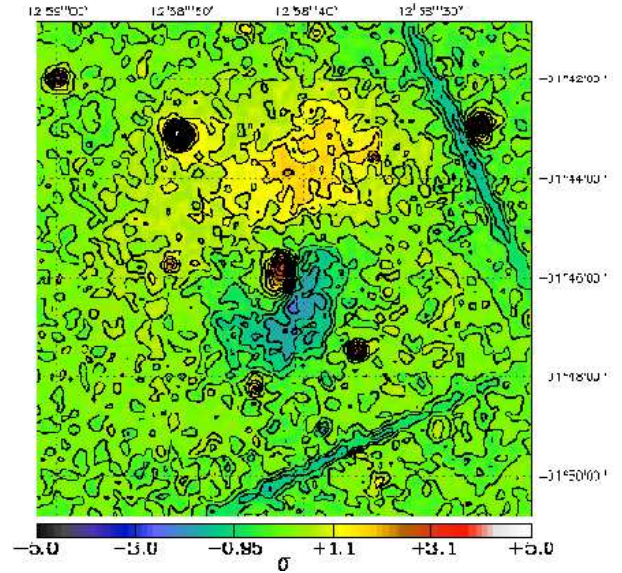}
  \end{center}
  \caption{Color-coded image in 0.8--10 keV with contours spaced by
 0.5 corresponds to the residuals after subtraction of the 2D elliptical
 $\beta$-model, smoothed with $4\farcs4$ Gaussian
 profile.}\label{fig:2dbetafit}
\end{figure}

The color-coded significance of the excess image with contours spaced by
0.5 after Gaussian smoothing with $\sigma=4\farcs4$, are shown in figure
\ref{fig:2dbetafit}. We can unambiguously confirm the excess emission
around the central cD galaxy and the excess of diffuse gas in the
direction to north. The image deficiency in south from the center is a
kind of artifact reflecting a steeper gradient of the surface brightness
distribution and compensating the excess in north shown in figures
\ref{fig:mos_opt} and \ref{fig:2dbetafit}, which depends on the position
of the image center and the angular range of the radial profile
fitting. This surface brightness distribution is similar to that of
A~2142 \citep{mar00}.

 In addition, with {\rm XMM-Newton}'s high angular resolution
 ($15\arcsec$ HPD) and good statistics, we could easily recognize three
 bright X-ray sources in the northern part, and resolve four point
 sources at radii of 2$\arcmin$--3$\arcmin$ from the center. Although
 some of them were identified as optical and radio sources, there were
 no measurements of their optical redshift. Therefore, we do not know
 whether these point sources are associated with the cluster or not.

\section{Spectro-Imaging Analysis}

\begin{figure}
  \begin{center}
    \FigureFile(80mm,80mm){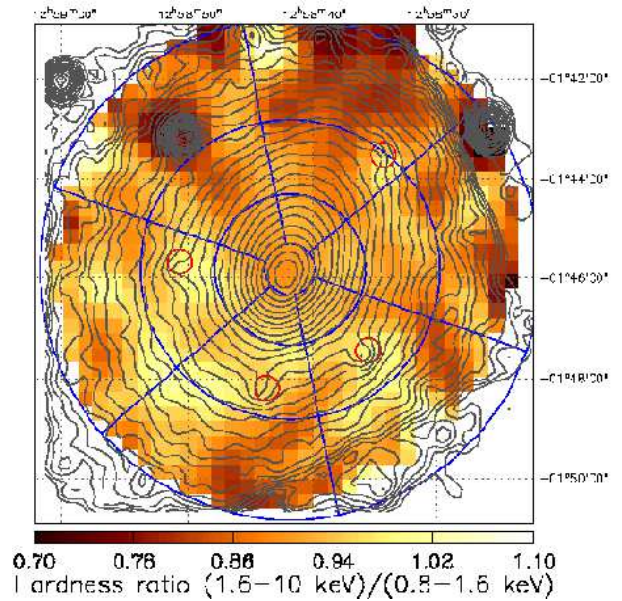}
  \end{center}
  \caption{Color-coded hardness ratio map overlaid on the X-ray surface
 brightness distribution, obtained by MOSs. The contours are
 logarithmically spaced by a factor of $1.20$, binned by $8\farcs8$
 pixels and filtered with a $\sigma=8\farcs8$ Gaussian profile in the
 energy band of 0.8--10 keV. The field is $10\arcmin\times 10\arcmin$
 with north up and east to the left. $HR$ map is binned by $17\farcs6$
 pixels and filtered with a $\sigma=21\farcs1$ Gaussian and shown in
 gray scale in the diameter of $10\arcmin$. The temperature and
 abundance for these 18 sectors are given in figure
 \ref{fig:sectorktab}. Red circles indicate point
 sources.}\label{fig:hrmap1}
\end{figure}

\begin{figure}
  \begin{center}
    \FigureFile(80mm,80mm){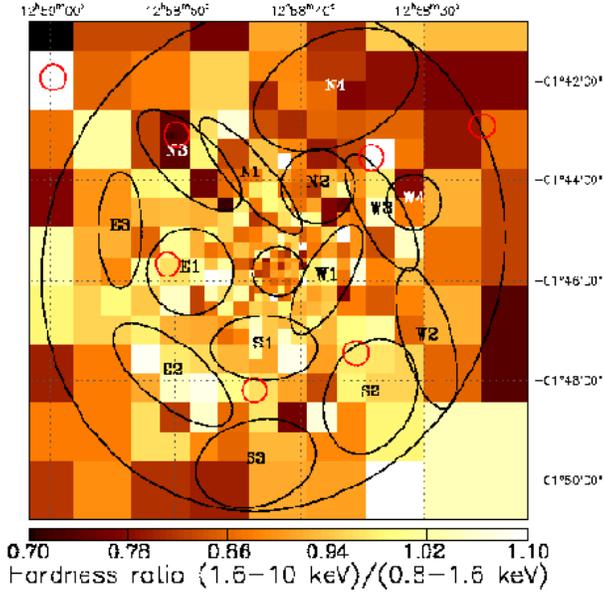}
  \end{center}
  \caption{Color-coded adaptively binned hardness ratio map with a
 fractional error of 0.10 in the same field as
 figure \ref{fig:hrmap1}. The encircled regions with an identification
 mark were selected for a spectral analysis. The positions of seven point
 sources in figure \ref{fig:hrmap1} are indicated by red
 circles.}\label{fig:hrmap2}
\end{figure}

We studied the spatially-resolved spectral feature using a hardness
ratio ($HR$) map. Since the {\rm XMM-Newton} telescope PSF has a very
weak energy dependence, we simply considered the spectral variation from
the X-ray hardness ratio. We defined $HR = I(1.6$--$10~{\rm
keV})/I(0.8$--$1.6~{\rm keV})$ as the ratio of the counts in the
respective energy bands, so that the counts in each energy band were
almost the same in order to minimize statistical error. The value of
$HR$ is not only a temperature indicator, but also an absorption
measure, and the mixture of complex spectral components. These
properties could be confirmed by a spectral analysis in the respective
regions referred to the $HR$ map. Based on a single-temperature
assumption, the relation between the $HR$ values and the
temperature($kT$) can be approximately expressed as $HR = 0.131 +
0.906\times (kT/5.62 - 0.387)^{0.294}$. $HR$ = 1 corresponds to $kT$ = 7
keV. Because the fractional errors of $HR$s are twice as large as those
of the surface brightness distribution in the 0.8--10 keV band, the
pixel sizes of the $HR$s were adjusted to obtain comparable statistics
to the surface-brightness distribution.

Figure \ref{fig:hrmap1} shows an $HR$ map with values in the range of
0.7--1.1 overlaid on X-ray surface brightness contours for a diameter of
$10\arcmin$ centered on the brightness peak by using MOSs data. These
images were binned by $8\farcs8$ pixels and filtered with a
$\sigma=13\farcs2$ Gaussian for the surface brightness contour and
$17\farcs6$ pixels and filtered with a $\sigma=21\farcs1$ Gaussian for
the $HR$ map. The systematically divided sector pattern is overlaid on
this map, which defines regions for another spectral analysis to compare
the $HR$ map.

Smaller bin sizes too much exaggerate any small-scale fluctuation in the
outer region, while larger ones smear out the significant structure in
the central region, and statistically clarify the structure in the outer
region. In order to show the overall feature in one figure, we used an
adaptive binning routine \citep{san01} to produce the $HR$ map. Although
the physical scale changed in this method, we ensured comparable
fractional errors per bin in each band. We defined the same $HR$ and
produced $HR$ with a fractional error of 0.10, as shown in figure
\ref{fig:hrmap2}. The minimum pixel size is $8\farcs8$ in the central
region and the maximum one is $1\farcm2$ in the outer region. The
overall feature is similar to the $HR$ map in figure
\ref{fig:hrmap1}. The encircled regions with an identification mark
(N1,~S1,$~\ldots$) were selected for the further spectral analysis
mentioned in subsection 5.3.

There exist significantly cool regions and a scalelength of a few 100
kpc, patchily distributed in the outer envelope. At larger scales, the
maps show that the hotter cluster gas lies southeast of its brightness
peak and cooler gas lies northwest, coinciding with the excess
structure, as shown in figure \ref{fig:2dbetafit}. The maps also shows
that the cluster brightness peak is moderately cool.


\section{Spectral Analysis}


\subsection{Overall Cluster Spectrum}

\begin{figure}
  \begin{center}
    \FigureFile(80mm,80mm){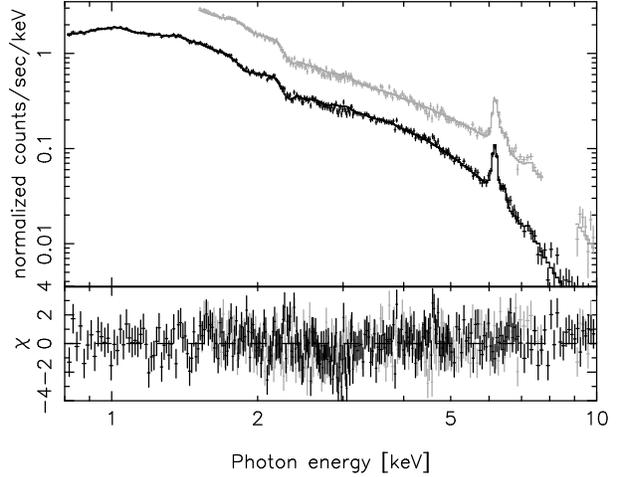}
  \end{center}
  \caption{Energy spectra of the MOSs (lower) and PN (upper) in the
 radius of $5\arcmin$. The solid lines indicate the best-fit isothermal
 model with $kT = 5.62 $ keV, an abundance of 0.36 times the solar
 value, and a redshift of 0.0801.}\label{fig:overallspec}
\end{figure}

The overall spectra of MOSs and PN were accumulated from all events
within a radius of $5\arcmin$, which included most of the cluster
emission and excluded obvious point sources. Each overall spectrum was
fitted with MEKAL, a collisionally ionized thermal plasma model
(\cite{mew85}, 1986; \cite{kaa92}; \cite{lie95}), incorporated with the
absorption of Galactic neutral gas. We excluded the energy bins around
the strong instrumental fluorescence lines of Ni, Cu, and Zn in the
energy range 7.8--9.0 keV from the PN spectral fitting. These lines were
not consistently subtracted from source spectra, since they were not
linearly related to the continuum of the particle-induced
background. All spectral fittings were performed using version 11.2.0 of
the XSPEC package.

The amount of absorption is fixed at a Galactic value of $N_H =
1.56\times 10^{20}~{\rm cm}^{-2}$ \citep{dic90}. The centroid energies
of the Fe-K${\rm \alpha}$ are systematically larger than the value
estimated from the optical redshift of 0.0845 \citep{str99}. Thus, the
redshift was also allowed to vary. We obtained the best-fit with a
redshift of $7.95^{+0.12}_{-0.11}\times 10^{-2}$ for MOSs and
$8.01^{+0.20}_{-0.17}\times 10^{-2}$ for PN. Since the MOSs and PN
spectra are consistent within the error bars, we adopted a value of
0.0801, derived from a simultaneous fit of MOSs and the PN spectra
as the redshift of A~1650.

There is also inconsistency for the temperature derived with the MOS and
PN data over the whole energy region of 0.3--10 keV. This can be due to
the soft component in the sky, noticed by \citet{nev03}, or an artifact
due to remaining calibration uncertainties. Therefore, we investigated
the influence of various low-energy cutoffs upon the temperature and
abundance for each instrument. It was found that there is an optimum
low-energy cutoff for each instrument, above which an even higher cutoff
energy will not significantly alter the resulting temperature. The
temperatures derived from MOSs and PN are consistent, and the converge
above the cutoff energy, which is 0.8 keV for the MOSs and 1.5 keV for
the PN data.

Consequently, we restricted our fit to the $0.8$--$10$ keV energy band
for MOSs and $1.5$--$10$ keV energy band for PN. The spectra were
corrected for vignetting and background subtracted, shown in figure
\ref{fig:overallspec}. The best-fit values of temperature ($kT$) and
abundance ($Z$) are $5.62^{+0.05}_{-0.07}$ keV and
$0.36^{+0.02}_{-0.01}$ solar for MOSs+PN respectively as listed in table
\ref{tab:results}. The observed flux in $2$--$10$ keV is $1.9 \times
10^{-11}$ erg s$^{-1}$ cm$^{-2}$, corresponding to a luminosity of
$5.6\times 10^{44}$ erg s$^{-1}$. We also analysed the ASCA GIS spectra
in the 0.7--10 keV band, and obtained $kT =5.84^{+0.19}_{-0.17}$ keV and
$Z = 0.36^{+0.06}_{-0.06}$ solar, which are consistent with the
above-mentioned values. We adopted the low-energy cutoffs thus derived
for the spatially-resolved spectral analysis discussed below, unless
otherwise stated.


\subsection{Radial and Azimuthal Profiles of Temperature and Abundance}

\begin{figure}
  \begin{center}
    \FigureFile(80mm,80mm){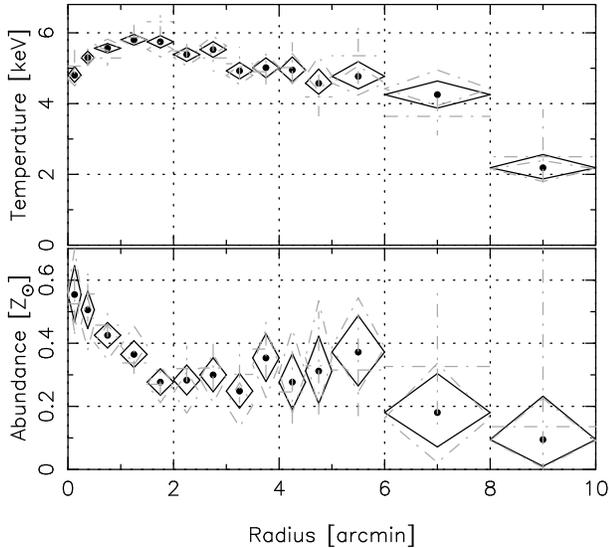}
  \end{center}
  \caption{Top: Radial temperature profile of A~1650 as a function
 of angular radius. Bottom: Radial abundance profile. The solid diamonds
 show the results of a combined fitting of MOSs and PN spectra. For a
 comparison, individual values of MOSs (dash-dotted diamonds) and PN
 (dash-dotted crosses) are also shown.}\label{fig:radial_kt_ab}
\end{figure}

We produced a radial temperature profile by excluding obvious point
sources and extracting spectra in annuli centered on the peak of the
X-ray emission. We extracted spectra in fourteen concentric annular
regions out to $10\arcmin$. Combined MOSs and PN spectra were binned to
at least 20 counts to allow the use of Gaussian statistics.

Each vignetting corrected and background-subtracted spectrum of MOS and
PN was then fitted simultaneously with an isothermal MEKAL model,
allowing the normalization, temperature and abundance to be free
parameters. In figure \ref{fig:radial_kt_ab} we show the radial profile
of the projected emission-weighted temperature and abundance. We can
recognize the obvious drop of temperature down to 4.8 keV in the
core. This spectral feature corresponds to the image excess at the
center shown in figure \ref{fig:2dbetafit}. Furthermore, temperatures
gradually decrease beyond about $2\arcmin$ and steeply drop down in
$8\arcmin$--$10\arcmin$.

Figure \ref{fig:radial_kt_ab} also shows the profile derived in the
conservative radial range where we have information from the Fe-K${\rm
\alpha}$ line. We observe 0.3 solar abundances outside of $r = 2\arcmin$
and a very sharp increase in the core to $0.55$ solar. The average
abundance is $0.36$, more typical of the value found by \citet{deg01}
for cooling flow clusters ($0.34\pm 0.01$). The profile exhibits a three
phase behavior, rather than a steadily declining profile. It decreases
up to $r = 3\arcmin$, and increases at a more or less constant rate with
a mean of 0.35 up to $r = 6\arcmin$ and decreases again.
\citet{irw01} showed that the abundance profiles of A~1795, A~2142, and
A~2163 clusters are also not monolithically declining with the radius.

\begin{figure}
  \begin{center}
    \FigureFile(80mm,80mm){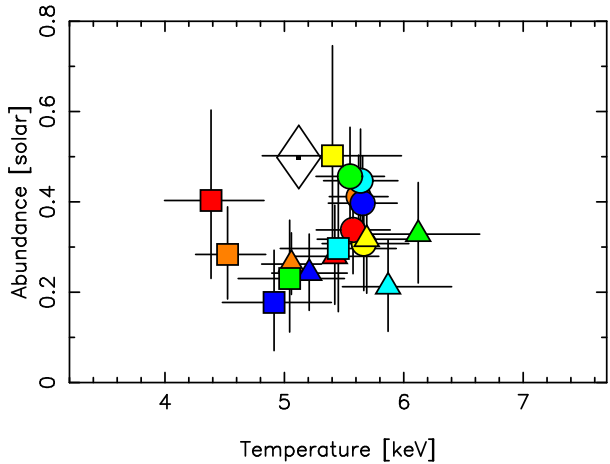}
  \end{center}
  \caption{Best-fit values of the temperature and abundance obtained
 by the MOSs
spectra in 19 systematically divided regions, shown in
figure \ref{fig:hrmap1}, as a sector pattern. The marker styles and
 colors indicate the radial positions, and position angles,
 respectively. Diamond: $r<0\farcm5$, filled circle: 6 sectors in $r= 0\farcm5$--$1\farcm5$, filled triangle: 6 sectors in
 1$\farcm$5--3$\arcmin$, filled square: 6 sectors in $r = 3\arcmin$--5$\arcmin$. Orange: -$50^\circ-10^\circ$ (north), red: $10^\circ-70^\circ$ (northeast), yellow: $70^\circ-130^\circ$ (southeast), green: $130^\circ-190^\circ$ (south), light blue: $190^\circ-250^\circ$ (southwest), blue: $250^\circ-310^\circ$ (northwest), where the angles are measured from north through east.}\label{fig:sectorktab}
\end{figure}

In order to further investigate the spatial distribution of the
temperature and abundance, we systematically divided the region within
$r = 5\arcmin$ into 19 regions with $r = 0\farcm5$, $1\farcm5$ and
$3\arcmin$ and 6 sectors in the azimuth except for $r < 0\farcm5$, as
shown in figure \ref{fig:hrmap1}. The same procedure of spectral fitting
was applied to each region. The results are shown concerning the
relation of the temperature and abundance in figure
\ref{fig:sectorktab}. It is noticed that at $r = 0\farcm5$--$1\farcm5$
the abundances azimuthally vary in the range of 0.31--0.46 solar at an
almost constant temperature of 5.6 keV, whereas at  $r =
1\farcm5$--3$\arcmin$, the temperatures increase up to 6.1 keV with
nearly constant abundances and at $r = 3\arcmin$--5$\arcmin$ the
temperatures are shifted down to  4.4 keV with a slight variation of the
abundances.

\subsection{Spectra in Characteristic Regions}

\begin{figure}
  \begin{center}
    \FigureFile(80mm,80mm){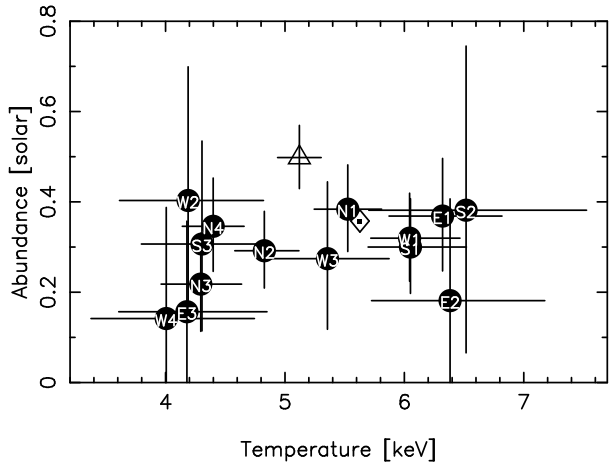}
  \end{center}
  \caption{Best-fit values of the temperature and abundance in selected
 regions with identification marks shown in figure
 \ref{fig:hrmap2}. Diamond, mean values in $r<5\arcmin$; triangle: in
 $r<0\farcm5$.}\label{fig:ktab}
\end{figure}

We chose 15 characteristic regions excluding obvious point sources, as
identified in figure \ref{fig:hrmap2}, in which the spectra were
analysed to quantify the statistical significance of the spectral
features in the $HR$ map. We performed a spectral fitting with a single
temperature MEKAL model.
The free parameters were the normalization, temperature, and abundance,
fixing $N_H$ to the Galactic value. The total counts used for the
spectral fitting in selected regions were in the range of
2000--12000/region, depending on the size of the regions and the radial
position angle on the surface-brightness distribution. Errors were
assigned with the 90\% confidence level for single parameters. The
best-fit values of the temperature and abundance thus obtained are
plotted in figure \ref{fig:ktab}, which are varied in the range of 4--6
keV, corresponding to $HR$ values of 0.78--1.0 and 0.15--0.55 solar,
respectively.

We find a temperature of $5.12$ keV and an abundance of 0.50 solar in
the central region ($r<0\farcm5$), as listed in table \ref{tab:results}
and shown in figure \ref{fig:ktab} with the triangle mark. The
calculated radiative cooling time is below the Hubble time in the
center. Therefore, we also fitted an absorbed thermal (MEKAL) plus
cooling flow (MKCFLOW) model to this region. The mass-deposition rate
within $r=0\farcm5$ is $3^{+5}_{-3}~M_\odot~{\rm yr}^{-1}$. Adding a
cooling flow component does not provide any improvement to the fit.

The lower values of temperature around 4 keV were obtained in regions of
N3,~N4,~E3,~S3,~W2, and W4. Region S2 shows the maximum value of the
temperature. Regions N1 and W3 are close to the average value. It is
obvious that the $HR$ values are well correlated to temperatures
obtained from the spectra. Referring to figure \ref{fig:ktab}, selected
regions are divided into 3 groups: high temperature group
(E1,~S1,~S2,~W1), low temperature and high abundance group (N4,~S3,~W2)
and low temperature and low abundance group (N3,~E3,~W4). The spectra in
these groups were fitted with the MEKAL model; the best-fit parameters
are listed in table \ref{tab:results}.

\begin{table*}
\begin{center}
\caption{Best-fit parameters of the spectral
 fitting$^*$. \label{tab:results}}
 \begin{tabular}{llcccccr}
  \hline\hline
  {Region}       & {Instrument}   & {$kT$}  & {$Z$}     & Redshift & {}             & {}   &  {}\\
  {}             & {}             & {(keV)} & {(solar)} & {($\times 10^{-2}$)}& {$\chi^2/$dof} & $HR$ & {Net counts\footnotemark[$^\dagger$]}\\
  \hline
  $r<5\arcmin$       & MOSs & $5.62^{+0.08}_{-0.07}$ &  $0.35^{+0.02}_{-0.02}$ & $7.95^{+0.12}_{-0.11}$ &  $328.4/272$ & $0.93$ & $182684$\\
  {}                 & PN       & $5.60^{+0.10}_{-0.07}$  &  $0.37^{+0.02}_{-0.02}$ & $8.01^{+0.20}_{-0.17}$ & $280.4/214$ & \ldots & $96459$\\
  {}                 & MOSs and PN & $5.62^{+0.05}_{-0.07}$ &  $0.36^{+0.02}_{-0.01}$ & $8.01^{+0.02}_{-0.02}$& $609.3/487$ & \ldots & $279143$\\
  $r<0\farcm5$       & MOSs & $5.12^{+0.18}_{-0.18}$ & $0.50^{+0.07}_{-0.07}$ & \ldots & $280.1/272$ & $0.89$ & $23062$\\
  {}                 & PN      & $5.13^{+0.26}_{-0.25}$ & $0.54^{+0.07}_{-0.07}$ & \ldots & $229.8/214$ & \ldots & $13106$\\
  {}                 & MOSs and PN & $5.14^{+0.14}_{-0.15}$ & $0.52^{+0.05}_{-0.05}$ & \ldots & $508.8/488$ & \ldots & $36168$\\
  N4+S3+W2           & MOSs & $4.45^{+0.21}_{-0.21}$ & $0.34^{+0.09}_{-0.08}$ & \ldots & $293.0/272$ & $0.85$ & $12634$\\
  N3+E3+W4           & MOSs & $4.55^{+0.37}_{-0.28}$ & $0.16^{+0.09}_{-0.08}$ & \ldots & $205.7/193$ & $0.84$ & $8068$\\
  E1+S1+S2+W1        & MOSs & $6.02^{+0.25}_{-0.19}$ & $0.32^{+0.05}_{-0.05}$ & \ldots & $318.3/272$ & $0.99$ & $36064$\\
  \hline
  \multicolumn{4}{@{}l@{}}{\hbox to 0pt{\parbox{180mm}{\footnotesize
      \par\noindent
      \footnotemark[$*$] PN data are partially masked out to avoid hot
  pixels near CCD boundary.
      \par\noindent
      \footnotemark[$\dagger$] Net counts are summed up in 0.8--10 keV for MOSs and 1.5--10 keV for PN.
    }\hss}}
  \end{tabular}
\end{center}
\end{table*}


\subsection{Galactic Diffuse Emission}

\begin{figure}
  \begin{center}
    \FigureFile(80mm,80mm){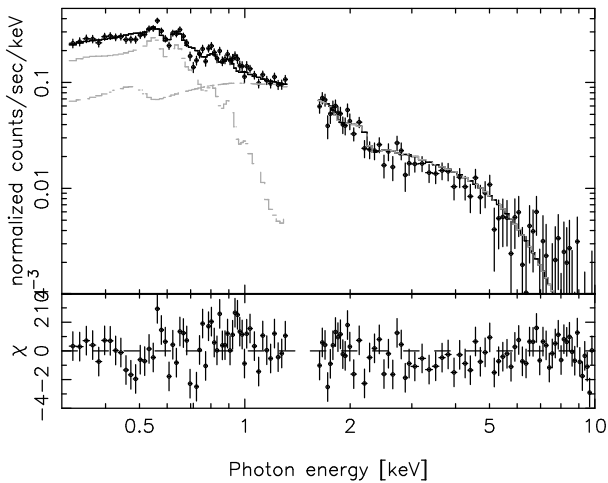}
  \end{center}
  \caption{Galactic diffuse and CXB spectrum derived from MOS data in
 the angular region of $r = 10\arcmin$--$15\arcmin$ free from the
 contribution of A~1650. The dashed line represents the MEKAL model for
 the Galactic thermal component, while the dash-dotted line represents
 the power-law model with $\Gamma = 1.4$ and $N_H=1.56\times 10^{20}~{\rm cm}^{-2}$ for CXB. The solid line is their combination.}\label{fig:diffuse_spec}
\end{figure}

The MOS spectrum of the outskirts of A~1650 in $r =
10\arcmin$--15$\arcmin$ was fitted with the model spectra of MEKAL and
the cosmic X-ray background. The MEKAL model corresponds to the Galactic
diffuse emission. The CXB model was fixed as a power-law (photon index
of 1.4) incorporated with the Galactic $N_H$ value in this direction. We
used data sets taken with the CLOSED filter position for background
subtraction, and excluded the fluorescent emission line of Al-K${\rm
\alpha}$ in the energy range of 1.35--1.6 keV from the fitting. The
result is shown in figure \ref{fig:diffuse_spec}. We can clearly
recognize strong emission lines of O\,{\footnotesize VII} (0.57 keV) and
O\,{\footnotesize VIII} (0.65 keV), which result in an oxygen abundance
of $0.099^{+0.010}_{-0.006}$ solar with a temperature of
$0.211^{+0.006}_{-0.008}$ keV. The depleted oxygen abundance is not
consistent with the previous picture of the hot interstellar medium
\citep{hay78}, which indicates that metallic elements are depleted by
trapping in dust grains, whereas the oxygen abundance is similar to the
solar value. The galactic diffuse soft X-ray spectrum could be explained
by a low-temperature component ($kT$ = 0.1 keV) from a local hot bubble
surrounding the solar system and extending to 100 pc or so, and a
high-temperature one ($kT$ = 0.2--0.3 keV) associated with radio loops
and other enhanced regions.  Using two temperature model, we tried to
fit the observed spectrum, since A~1650 is located in the high galactic
latitude and in the direction of a part of Loop I \citep{sno95}. We
obtained a slightly better fitting with $kT$ = 0.11 and 0.24 keV and
abundance of 0.9 solar, though the coupling of parameters was rather
strong. This problem should be further investigated using data obtained with {\rm XMM-Newton} observations.

\section{Point Sources}

\begin{figure}
  \begin{center}
    \FigureFile(80mm,80mm){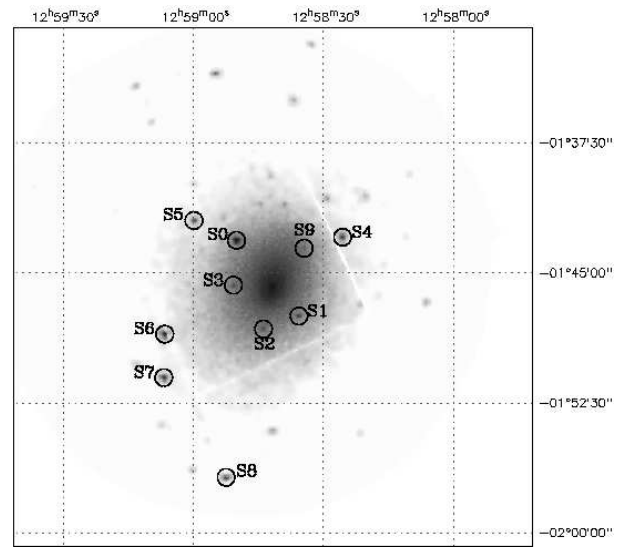}
  \end{center}
  \caption{Adaptively smoothed broadband (0.3--10 keV) MOSs image of the
 $30\arcmin\times 30\arcmin$ region surrounding the center of
 A~1650. The open circles correspond to the bright point sources given
 in table \ref{tab:psresults}.}\label{fig:MOS_psid}
\end{figure}

\begin{figure}
  \begin{center}
    \FigureFile(80mm,80mm){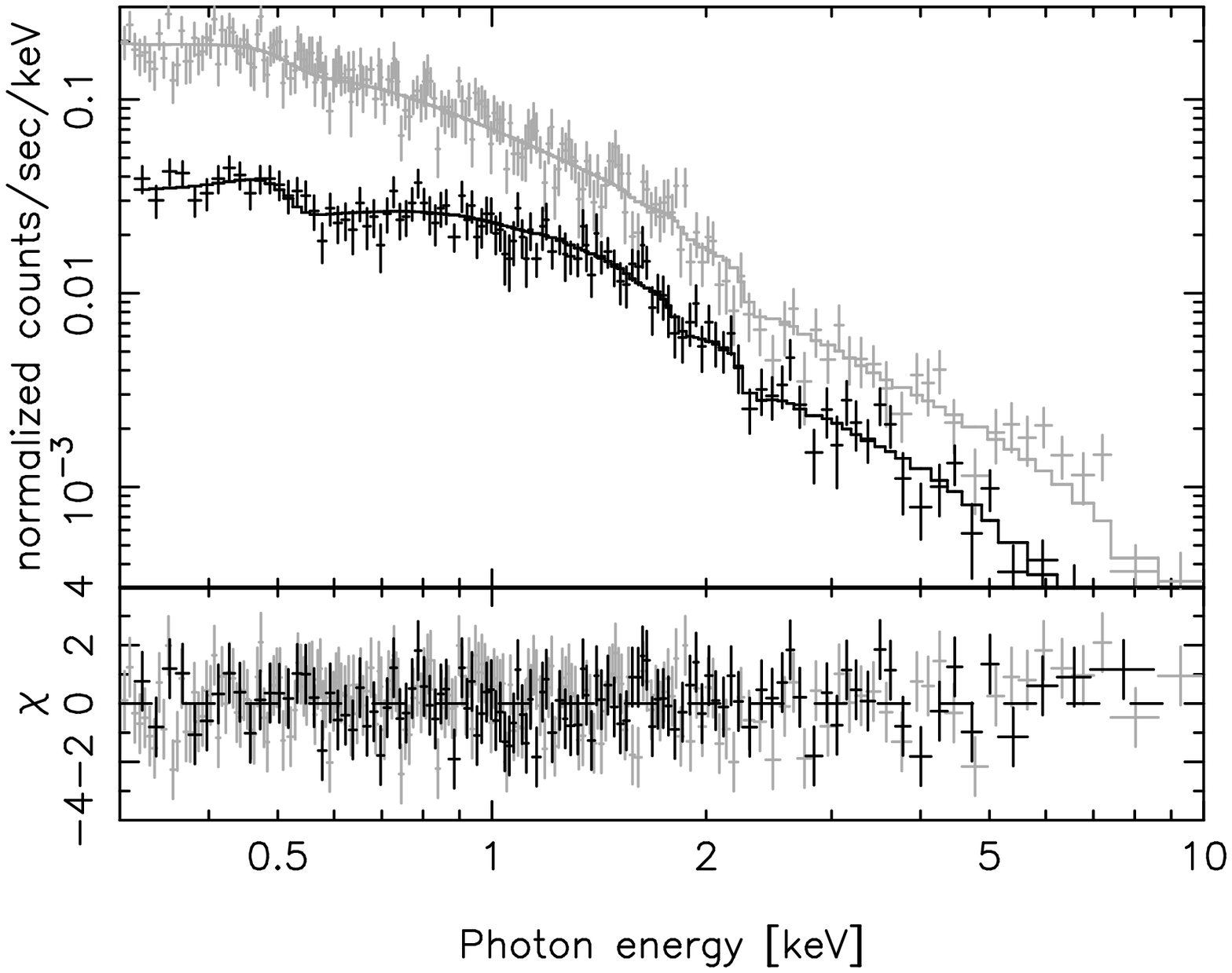}
    \FigureFile(80mm,80mm){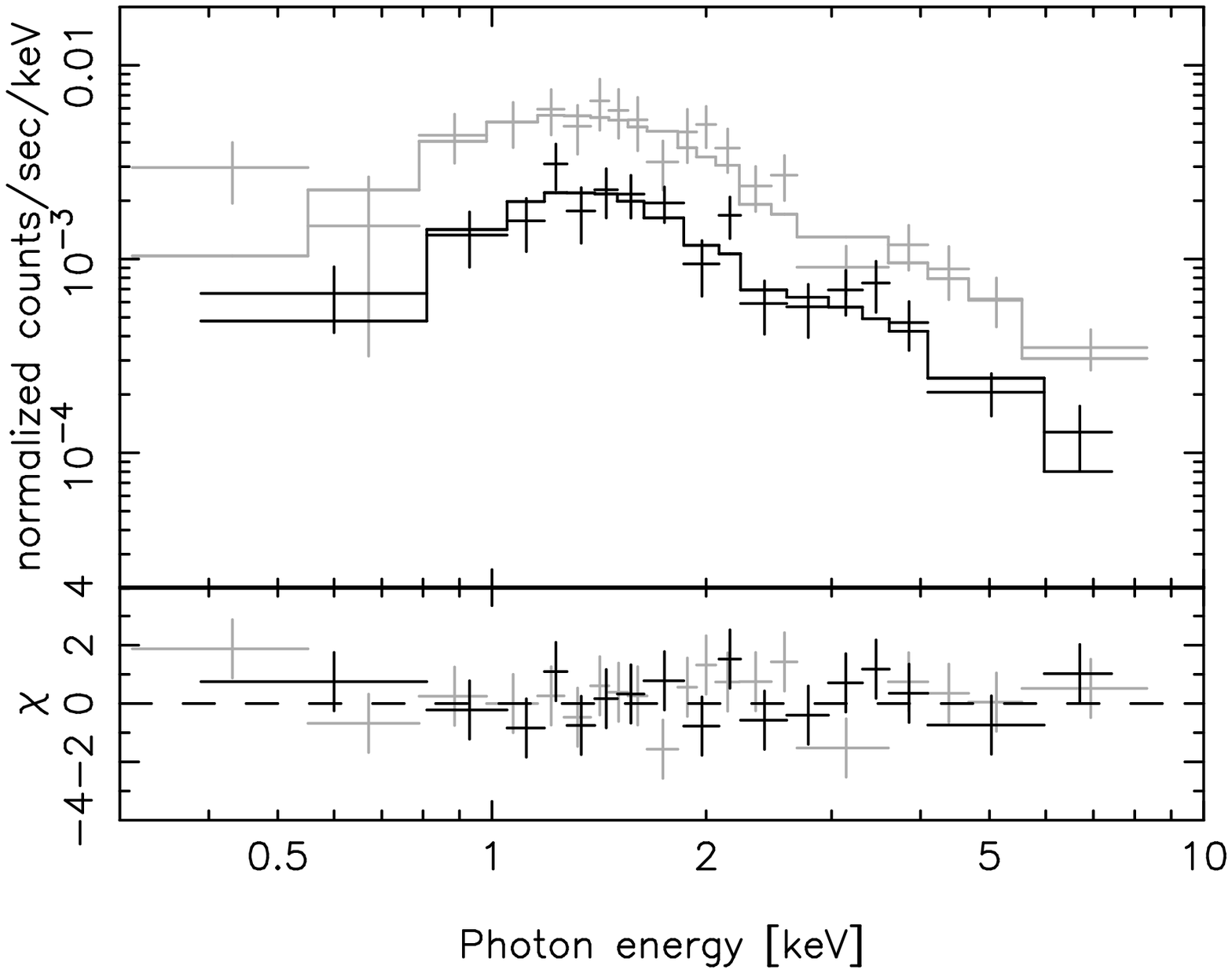}
  \end{center}
  \caption{Top: MOSs (lower) and PN (upper) spectra of S0 with the
 best-fit model (a power-law with $\Gamma = 2.15^{+0.07}_{-0.04}$
 modified by the Galactic absorption). Bottom: Those of S5 with the
 best-fit intrinsic absorption model (an absorbed power-law with $\Gamma = 1.72^{+0.17}_{-0.30}$ and $N_H = 4.14^{+1.49}_{-1.77}\times 10^{21}~{\rm cm}^{-2}$)}\label{fig:XMM_psspec}
\end{figure}

Unexpectedly, many sources were observed around A~1650, which were not
found in the ASCA X-ray image. Ten bright point sources shown in figure
\ref{fig:MOS_psid} are listed in table \ref{tab:psresults} with
coordinates, net counts, spectral indices, $N_H$ values,
$\chi^2_\nu$/dof and fluxes in the 2--10 keV band. Source spectra were
extracted from circular regions of $0\farcm25$ radius for MOSs and
PN. Background spectra were taken from identical circular source-free
regions in the vicinity of each source. We fit the model to all three
datasets between 0.3 and 10 keV simultaneously, and the spectra were
grouped to have a minimum of 20 counts in each bin. Their spectra,
subtracting the diffuse cluster component, were fitted by a power-law
model, except for S8, since significant emission lines were not
recognized in their spectra. The best-fits to the spectra of S0 and S5
are shown in figure \ref{fig:XMM_psspec}, demonstrating that S5 requires
substantial amounts of excess absorption.

S0 is the brightest object, and is located at a radius of $3\farcm35$
from the cluster center. The peak intensity is almost comparable to that
of the cluster center. This angular separation and intensity make it
possible to detect with the ASCA observation. This means that S0 would
be highly variable. The 0.3--10 keV spectrum is featureless and well
fitted by a power-law with a photon index of $2.15^{+0.07}_{-0.04}$
($\chi^2_\nu$/dof = 289.0/305), as shown in figure
\ref{fig:XMM_psspec}. A thermal plasma model fits it poorly. The
spectrum of S0 is similar to the steep spectra found in BL Lac objects
(\cite{cil95}). S1, S2, S3, and S9 are located within a radius of
2$\arcmin$--3$\arcmin$. Since the spectra of AGNs are rather
homogeneously represented by a power-law with a photon index of 1.7--2.0
\citep{nan94}, S1, S2, S3, and S9 may be AGNs. S5 shows the strong
intrinsic absorption caused by the surrounding matter of the source. The
intrinsic absorption inferred from spectral fits is
$4.14^{+1.49}_{-1.77}$ $\times 10^{21}~{\rm cm}^{-2}$ and the photon
index is $1.72^{+0.17}_{-0.30}$. This amount of absorption and the
photon index imply that S5 is an obscured AGN. S7 is a bright radio
source (PKS 1256$-$015). S8 is identified with a star (HD~112719), and
shows a thermal spectrum. If Fe-K${\rm \alpha}$ lines could be found in
the spectra, we could estimate the redshift. Because optical
identification of these objects is rather poor, we have to investigate
them further.

\begin{table*}
\begin{center}
\caption{Characteristics of point sources. \label{tab:psresults}}
\small{
 \begin{tabular}{lccrcccrl}
  \hline\hline
  {} & R.A.      & Decl.      & \multicolumn{1}{c}{Net}    & {}       & $N_H$
     & {}               & \multicolumn{1}{c}{$F_X$(2--10 keV)}        & {}\\
  ID & (J2000.0) & (J2000.0)  & counts$^*$ & $\Gamma^\dagger$ & ($\times 10^{21}{\rm cm}^{-2}$)
     & $\chi_\nu^2$/dof & (erg s$^{-1}$cm$^{-2}$) & Comments\\
  \hline
S0 & 12~58~49.92 & -01~43~04.5 & 7729 & $2.15^{+0.07}_{-0.04}$ &
  $0.00^{+0.10}_{-0.00}$ & 289.0/305 & 21.09$\times 10^{-14}$ &  \\
S1 & 12~58~35.52 & -01~47~26.4 & 1433 & $1.93^{+0.14}_{-0.09}$ & $0.00^{+0.14}_{-0.00}$ & 100.6/96  & 10.66$\times 10^{-14}$ &  \\
S2 & 12~58~43.68 & -01~48~11.4 &  677 & $2.28^{+0.69}_{-0.57}$ & $2.01^{+2.33}_{-1.53}$ & 116.5/100 &  3.19$\times 10^{-14}$ &  \\
S3 & 12~58~50.64 & -01~45~39.9 &  672 & $1.82^{+0.37}_{-0.21}$ & $0.00^{+0.42}_{-0.00}$ & 106.7/101 &  3.30$\times 10^{-14}$ &  \\
S4 & 12~58~25.44 & -01~42~53.6 & 1395 & $2.11^{+0.26}_{-0.14}$ & $0.20^{+0.46}_{-0.20}$ &  64.7/76  &  6.22$\times 10^{-14}$ &  \\
S5 & 12~58~59.76 & -01~41~55.9 &  727 & $1.72^{+0.17}_{-0.30}$ & $4.14^{+1.49}_{-1.77}$ &  46.6/43  &  7.99$\times 10^{-14}$ & EXSS~1256.5-0134 \\
S6 & 12~59~06.48 & -01~48~29.3 & 3440 & $1.83^{+0.10}_{-0.08}$ & $0.31^{+0.19}_{-0.16}$ & 166.0/160 & 22.06$\times 10^{-14}$ &  \\
S7 & 12~59~06.72 & -01~50~59.5 & 1941 & $1.77^{+0.20}_{-0.09}$ & $0.15^{+0.45}_{-0.14}$ &  92.3/93  & 14.74$\times 10^{-14}$ & PKS~1256-015\\
S8 & 12~58~52.32 & -01~56~45.0 & 1252 &
  \multicolumn{2}{c}{$0.53^{+0.05}_{-0.03}$ keV, $0.12^{+0.03}_{-0.03}Z_\odot$} & 60.3/54 & 0.24$\times 10^{-14}$ & HD~112719\\
S9 & 12~58~34.32 & -01~43~31.7 &  590 & $1.50^{+0.59}_{-0.22}$ &  $0.38^{+0.16}_{-0.38}$ &  70.3/84  &  4.97$\times 10^{-14}$ & \\
 \hline
  \multicolumn{4}{@{}l@{}}{\hbox to 0pt{\parbox{180mm}{\footnotesize
      \par\noindent
      \footnotemark[$*$] {Net counts in $r< 0.25\arcmin$ from the center
  of each source and in 0.3--10keV.}
      \par\noindent
      \footnotemark[$\dagger$] {Results of the power-law fits except for S8
  which is fitted with MEKAL model.}
    }\hss}}
  \end{tabular}
  }
\end{center}

\end{table*}


\section{Discussion}

The overall spectrum of A~1650 is well-expressed with a
single-temperature model, which indicates that this cluster is
isothermal and relaxed. It is not easy to find the multi temperature
components only from a spectral fitting. The spectral analysis referring
to the $HR$ map makes it possible to spatially distinguish several
temperature components of 4--6 keV and the 0.2--0.5 solar abundance from
the overall spectrum. This fact is also confirmed by a spectral analysis
in the systematically sub-divided regions. The $HR$ map more clearly
shows the distribution of temperature components, as shown in figure
\ref{fig:ktab}.

There exist significantly cool regions with a temperature of around 4
keV and a scalelength of a few 100 kpc patchily distributed in the outer
envelope. Also, somewhat hot regions with a temperature of around 6 keV
are extended in the southeast direction. Some of point sources are
located in the hot region. The central region within a radius of
$0\farcm5$ obviously shows a lower temperature and a higher abundance
compared to the average values within a radius of 5$\arcmin$, which
seems to be related to the existence of a cD galaxy. The image excess
region extended to the north in figure \ref{fig:2dbetafit} shows a lower
temperature of 5 keV.

The characteristic time, $\delta\tau$, required to smear out the electron
temperature gradient and arising from the effect of the thermal
conduction alone would be $\delta\tau = \delta r/\bar{v} = \delta
r/(\frac{2}{3}\frac{\kappa}{n_ekT_e}\frac{d(kT_e)}{dr})$
\citep{spi62}. We adopted typical values of $n_e=5.0\times 10^{-3}~{\rm
cm}^{-3}$ and $\Delta T_e=1.0$ keV. Assuming that the scalelength,
$\delta r$, of this temperature gradient is 50 kpc, the diffusion
timescale, $\delta\tau$, is estimated to be $3\times 10^8$ yr, shorter
than the expected age of the cluster. We discuss several possible
explanations for this temperature structure.

We could see no evidence for a large-scale temperature variation and
X-ray surface brightness irregularity, as seen in the Coma and Ophiuchus
clusters (Watanabe et al. 1999, 2001); moreover, X-ray emissions are more
concentrated in the center. This means that this cluster has not
experienced recent merger activity. Therefore, direct shock heating via
a major merger does not account for this structure. It is partly
plausible that hot regions associated with point sources would be caused
by the energy output of the AGN activity.

On the other hand, cool regions could be explained by the contribution of
infalling groups of galaxies not yet thermalized. In the outer envelope
the contribution of this component is significant to decrease the
temperature of the main body of a cluster. On the contrary, the large
difference in the temperature is not significantly recognized in the
central region. In this case, a somewhat higher temperature in the
boundary region is caused by adiabatic compression of the surrounding
intracluster gas when the groups of galaxies infall to the main body of
a cluster.

There is always a large cD galaxy at the center of massive cooling flow
clusters, and around 71\% of all cD galaxies in cooling flows show
evidence of radio activity \citep{bur90}. \citet{boh02} finds that the
power of the AGN jets from radio sources is more than sufficient to heat
the cooling flows and to reduce the mass-deposition rates. In the case
of A~1650, because it hosts a radio-quiet cD galaxy and the X-ray
emission is more peaked, we expect a massive cooling flow. While the
temperature of the ICM slightly drops even in the central $0\farcm25$ of
the cluster, the addition of a cooling flow component does not improve
the fits. Thus, A~1650 bears no evidence of multiphase gas, refuting an
earlier imaging study of $\dot{M} \sim 200~ M_\odot$ yr$^{-1}$
\citep{per98}. We suggest that adiabatic heating via infalling of groups
is sufficient to heat the cooling flow and to reduce the mass-deposition
rate.


\section{Conclusion}

The non-uniform distribution of the temperature and abundance in an
angular extent of $10\arcmin$ (1.4 Mpc) in A~1650 was obtained by a
spectral analysis in 19 systematically-subdivided regions and in the
specified regions referred to the hardness ratio ($HR$) map with bin
sizes of $17\farcs6$. The value of $HR$ is a useful indicator to
investigate the spatial variation of the spectral feature. The average
temperature and abundance were obtained to be $5.62^{+0.05}_{-0.07}$ keV
and $0.36^{+0.02}_{-0.01}$ solar within a radius of $5\arcmin$,
respectively, where the redshift was derived to be
$8.01^{+0.02}_{-0.02}\times 10^{-2}$ against an optical value of
$8.45\times 10^{-2}$. It turned out that cool and hot regions with a
temperature of 4--6 keV and a scalelength of a few 100 kpc are patchily
distributed in the outer envelope, whereas the radially increasing
temperature and decreasing abundance are significantly observed in the
central core region. This temperature structure could be explained by
the infalling group of galaxies to the main body of the cluster. In the
outskirts of A~1650 the spectrum of the Galactic diffuse emissions was
obtained with significant detection of O\,{\footnotesize VII} and
O\,{\footnotesize VIII} lines. Moreover, this cluster extends to a
radius of $10\arcmin$, where we found more than 30 point sources. Their
association with the cluster is still an open question, since their
distances are not well-defined.

\bigskip
\bigskip
We thank an anonymous referee for a careful reading of the manuscript
and helpful comments. This work was based on observations obtained with
{\rm XMM-Newton}, an ESA science mission with instruments and
contributions directly funded by ESA member states and USA (NASA). This
work was supported in part by a Grant-in-Aid for Specially Promoted
Research contract 07102007, from the Ministry of Education, Culture,
Sports, Science and Technology.



\begin{thebibliography}{}
\bibitem[Abell et al.(1989)]{abe89} Abell, G. O., Corwin, H. G., Jr., \& Olowin, R. P. 1989, \apjs, 70, 1
\bibitem[Anders and Grevesse(1989)]{and89} Anders, E., \& Grevesse,
										N. 1989,
										Geochim. Cosmochim. Acta, 53,
										197
\bibitem[Arnaud et al.(2001)]{arn01} Arnaud, M., Neumann, D. M.,
										Aghanim, N., Gastaud, R.,
										Majerowicz, S., \& Hughes,
										J. P. 2001, \aap, 365, L80
\bibitem[Bagchi, Kapahi(1994)]{bag94} Bagchi, J., \& Kapahi, V. K. 1994,
							J. Astrophys. Astr., 15, 275
\bibitem[B\"ohringer et al.(2002)]{boh02} B\"ohringer, H., Matsushita,
							K., Churazov, E., Ikebe, Y., \& Chen, Y. 2002, \aap, 382, 804
\bibitem[Burns(1990)]{bur90} Burns, J. O. 1990, \aj, 99, 14
\bibitem[Ciliegi et al.(1995))]{cil95} Ciliegi, P.,
										Bassani, L., \& Caroli, E. 1995,
										\apj, 439, 80
\bibitem[De Grandi and Molendi(2001)]{deg01} De Grandi, S., \& Molendi, S. 2001, \apj, 551, 153
\bibitem[Dickey, Lockman(1990)]{dic90} Dickey, J. M., \&  Lockman, F. J. 1990, \araa, 28, 215
\bibitem[Furusho et al.(2001)]{fur01} Furusho, T., Yamasaki, N. Y., Ohashi, T., Shibata, R., \& Ezawa, H. 2001, \apj, 561, L165
\bibitem[Hayakawa et al.(1978)]{hay78} Hayakawa, S., Kato, T., Nagase,
										F., Yamashita, K., \& Tanaka, Y., 1978, \aap, 62, 21
\bibitem[Ikebe et al.(2002)]{ike02} Ikebe, Y., Reiprich, T. H.,
										B\"ohringer, H., Tanaka, Y., \&
										Kitayama, T. 2002, \aap, 383, 773
\bibitem[Irwin and Bregman(2001)]{irw01} Irwin, J. A., \& Bregman,
										J. N. 2001, \apj, 546, 150
\bibitem[Jansen et al.(2001)]{jan01} Jansen, F., \etal\ 2001, \aap, 365, L1
\bibitem[Kaastra(1992)]{kaa92} Kaastra, J. S. 1992, An X-Ray Spectral Code for Optically Thin Plasmas (Internal SRON-Leiden Report, updated version 2.0)
\bibitem[Kaastra et al.(2001)]{kaa01} Kaastra, J. S., Ferrigno, C., Tamura, T., Paerels, F. B. S., Peterson, J. R., \& Mittaz, J. P. D. 2001, \aap, 365, L99
\bibitem[Liedahl et al.(1995)]{lie95} Liedahl, D. A., Osterheld, A. L., \& Goldstein, W. H. 1995, \apj, 438, L115
\bibitem[Markevitch et al.(2000)]{mar00} Markevitch, M., \etal\ 2000,
							\apj, 541, 542
\bibitem[Markevitch et al.(1998)]{mar98} Markevitch, M., Forman, W. R., Sarazin, C. L., \& Vikhlinin, A. 1998, \apj, 503, 77
\bibitem[Mewe et al.(1985)]{mew85} Mewe, R.,  Gronenschild, E. H. B. M., \& van den Oord, G. H. J. 1985, \aaps, 62, 197
\bibitem[Mewe, Lemen, \& van den Oord(1986)]{mew86} Mewe, R., Lemen, J. R., \& van den Oord, G. H. J. 1986, \aaps, 65, 511
\bibitem[Nandra, Pounds(1994)]{nan94} Nandra, K., \& Pounds, K. A.  1994, \mnras, 268, 405
\bibitem[Nevalainen et al.(2003)]{nev03} Nevalainen, J., Lieu, R., Bonamente, M., \& Lumb, D. 2003, \apj, 584, 716
\bibitem[Peres et al.(1998)]{per98} Peres, C. B., Fabian, A. C., Edge, A. C., Allen, S. W., Johnstone, R. M., \& White, D. A. 1998, \mnras, 298, 416
\bibitem[Ricker(1998)]{ric98} Ricker, P. M. 1998, \apj, 496, 670
\bibitem[Roettiger, Flores(2000)]{roe00} Roettiger, K., \& Flores,
							R. 2000, \apj, 538, 92
\bibitem[Roettiger et al.(1997)]{roe97} Roettiger, K., Loken, C., \& Burns, J. O. 1997, \apjs, 109, 307
\bibitem[Sanders, Fabian(2001)]{san01} Sanders, J. S., \& Fabian, A. C. 2001, \mnras, 325, 178
\bibitem[Sanders, Fabian(2002)]{san02} Sanders, J. S., \& Fabian, A. C. 2002, \mnras, 331, 273
\bibitem[Snowden et al.(1995)]{sno95} Snowden, S. L., \etal\ 1995, \apj, 454, 643
\bibitem[Spitzer(1962)]{spi62} Spitzer, L., Jr. 1962, Physics of Fully Ionized Gases~(New York:Interscience)
\bibitem[Struble, Rood(1999)]{str99} Struble, M. F., \& Rood, H. J.,
							1999, \apjs, 125, 35
\bibitem[Str\"uder et al.(2001)]{str01} Str\"uder, L., \etal\ 2001, \aap, 365, L18
\bibitem[Tamura et al.(2001)]{tam01} Tamura, T., \etal\ 2001, \aap, 365, L87
\bibitem[Turner et al.(2001)]{tur01} Turner, M. J. L., \etal\ 2001, \aap, 365, L27
\bibitem[Watanabe et al.(2001)]{wat01} Watanabe, M., Yamashita, K.,
							Furuzawa, A., Kunieda, H., \& Tawara,
							Y. 2001, \pasj, 53, 605
\bibitem[Watanabe et al.(1999)]{wat99} Watanabe, M., Yamashita, K., Furuzawa, A., Kunieda, H., Tawara, Y., \& Honda, H. 1999, \apj, 527, 80
\bibitem[White(2000)]{whi00} White, D. A. 2000, \mnras, 312, 663
\bibitem[White et al.(1997)]{whi97} White, D. A., Jones, C., \& Forman, W. 1997, \mnras, 292, 419

\end{thebibliography}
\end{document}